\documentclass[final]{cvpr}

\usepackage{times}
\usepackage{epsfig}
\usepackage{graphicx}
\usepackage{amsmath}
\usepackage{amssymb}

\usepackage{mathtools}
\usepackage{caption}
\usepackage[pagebackref=true,breaklinks=true,colorlinks,bookmarks=false]{hyperref}

\begin{document}

\title{Multi-Instrumentalist Net: Unsupervised Generation of Music from Body Movements}


\author{%
  Kun Su\\
  Department of Electrical \& Computer Engineering\\
  University of Washington, Seattle, WA 98195
  \and
  Xiulong Liu \\
  Department of Electrical \& Computer Engineering  \\
  University of Washington, Seattle, WA 98195
  \and
  Eli Shlizerman \\ 
  Department of Applied Mathematics \\
  Department of Electrical \& Computer Engineering \\  University of Washington, Seattle, WA 98195\\
}

\twocolumn[{
\renewcommand\twocolumn[1][]{#1}
\maketitle
\vspace{-0.55in}
\begin{center}
    \includegraphics[width=0.75\linewidth]{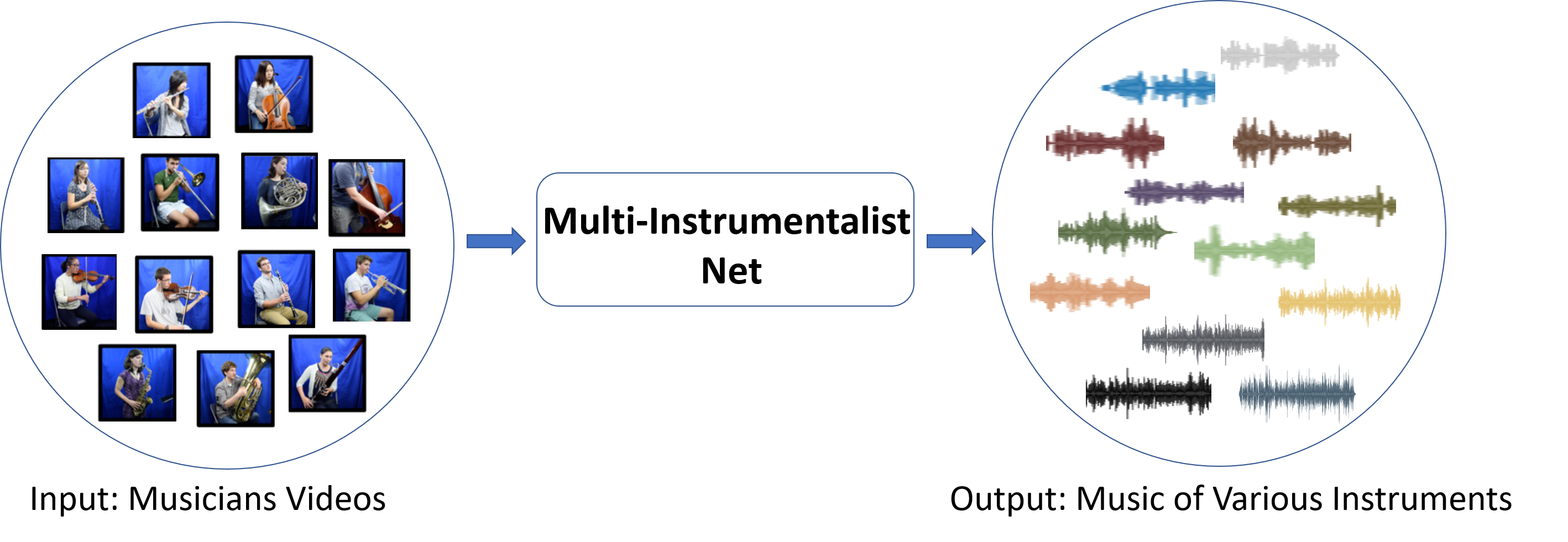}
    \vspace{-0.15in}
    \captionof{figure}{Multi-instrumentalist Net: Unsupervised music generation from body movements. Given an input of video frames of a musician playing a musical instrument, Multi-instrumentalist Net (MI Net) generates the associated music for it. Please see supplementary video and supplementary generated music samples.}\label{fig:teaser}
\end{center}
}]
\begin{abstract}
We propose a novel system that takes as an input body movements of a musician playing a musical instrument and generates music in an unsupervised setting. Learning to generate multi-instrumental music from videos without labeling the instruments is a challenging problem. To achieve the transformation, we built a pipeline named 'Multi-instrumentalistNet' (MI Net). At its base, the pipeline learns a discrete latent representation of various instruments music from log-spectrogram using a Vector Quantized Variational Autoencoder (VQ-VAE) with multi-band residual blocks. The pipeline is then trained along with an autoregressive prior conditioned on the musician's body keypoints movements encoded by a recurrent neural network. Joint training of the prior with the body movements encoder succeeds in disentanglement of the music into latent features indicating the musical components and the instrumental features. The latent space results in distributions that are clustered into distinct instruments from which new music can be generated.
Furthermore, the VQ-VAE architecture supports detailed music generation with additional conditioning. We show that a Midi can further condition the latent space such that the pipeline will generate the exact content of the music being played by the instrument in the video. We evaluate MI Net on two datasets containing videos of 13 instruments and obtain generated music of reasonable audio quality, easily associated with the corresponding instrument, and consistent with the music audio content.
\end{abstract}
\section{Introduction}
\begin{figure*}[!t]
\begin{center}
    \includegraphics[width=0.77\textwidth]{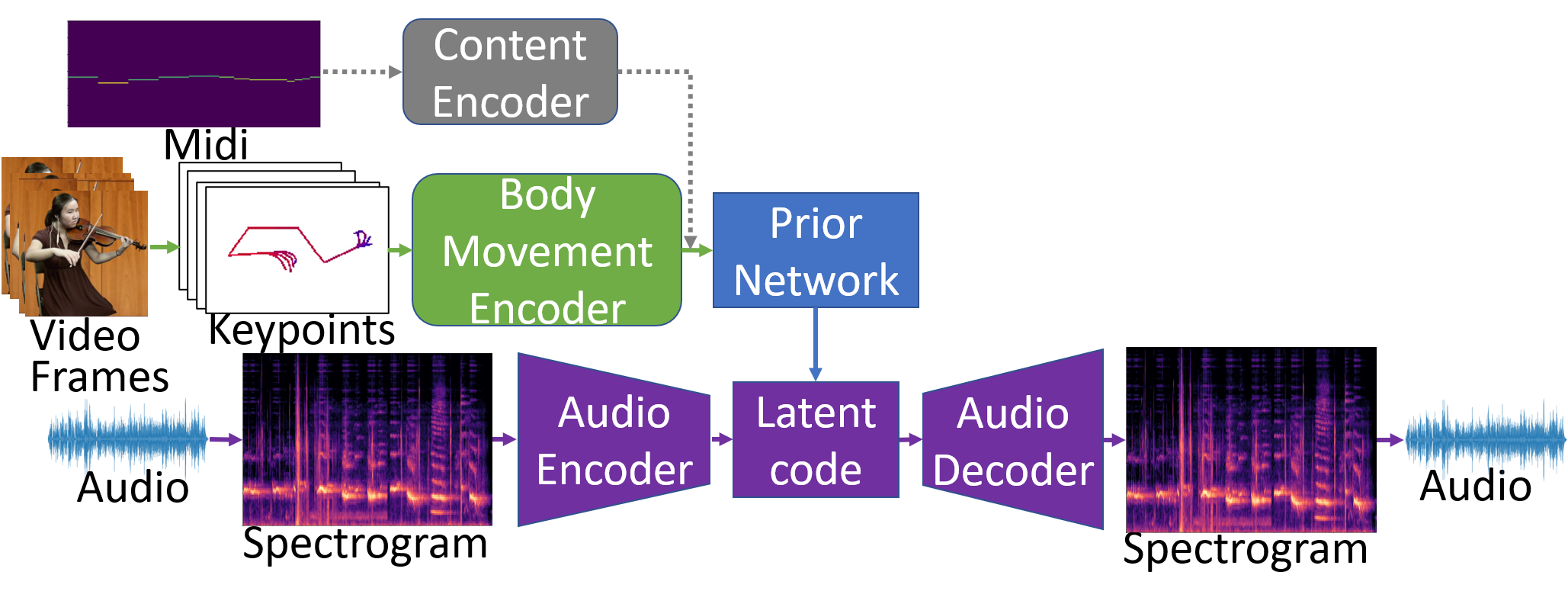}
    \caption{System overview of MI Net. VQ-VAE (bottom) is used to reconstruct audio sequences of various instruments and infers a latent representation of music (latent code). VQ-VAE is conditioned by a prior network (middle) that encodes body movements w/wo MIDI content. Further, given an input of Video Frames of musician playing the instrument, Multi-instrumentalist Net (MI Net) generates the music for that instrument.}
    \label{fig:overview}
\end{center}
\end{figure*}

\begin{flushright}
\textit{``This is what it sounds like"\\ When Doves Cry, Purple Rain, Prince
}\end{flushright}
A multi-instrumentalist is a musician who plays two or more musical instruments and easily transitions from one instrument to another. For example, the famous multi-instrumentalist, Prince, played all of the 27 instruments featured in his first album `For You'. Such a talent is expressed in the ability to disentangle the uniqueness of each instrument along with maneuvering the similarities across instruments.
\\
Music theory defines an assortment of components, such as pitch, rhythm, dynamics, timbre, that characterize the music. The combination and variation of these components creates numerous and substantially different types of music. One of the fundamental factors in the distinction between music instruments is timbre. While  timbre is the dominant component in the association of music with an instrument, it is not the sole one. Such associations appear to be tangled in evident by perceptual experiments suggesting that non-professionals would not be able to tell which instrument is playing from just listening to a piece of single instrument music. Such a complexity exists in a computational setting as well, where disentanglement of timbre from audio is not a straightforward task.
\\
In a situation where the audio signal could be ambiguous, the visual information, i.e., a video of the musician playing, greatly simplifies such associations. In perception, visual information helps the brain to disentangle the source of the sounds. For example, the association of the type of an instrument that plays the music becomes a simple task. This is supported by recent computational research which shows that visual information has the potential to significantly enhance audio tasks, such as the sound separation.
\\
Indeed visual cues along with sound information complete each other. However, is it possible to step even further and to set a computational system to generate instrumental music from visual cues alone? It turns out to be a challenging problem, but has the potential to identify the components that are involved in 
generating music. In recent years, methods employing deep-learning techniques have shown plausibility of accomplishing such a transformation between visual cues and music. Current methods have achieved convincing music generation results from visual information, such as body keypoints, or full video. However, they still incorporate limitations in terms of generating music for several instruments. These limitations are expressed in relying on strong supervision that requires numerous examples accompanied by instrument labels or an implementation of training of a distinct model per each instrument.
\\
In this work, we develop a system which is a single model that generates different types of instrumental music from unlabeled videos. In particular, we introduce `Multi-instrumentalist Net' (MI Net) that succeeds to generate in an unsupervised manner various instrumental music signals from videos of musicians playing music. An overview of the system is shown in Fig. \ref{fig:overview}. At the heart of our system is a  Vector-Quantized Variational Autoencoder (VQ-VAE) network~\cite{van2017neural} that learns in an unsupervised way a discrete latent space for audio features by reconstructing the audio input. We then simultaneously train a prior network of the latent space with a visual information encoder network. In particular, we use body keypoints as our visual features. This step turns out to be critical in finding the correlations between the visual information and the audio features captured in the latent space. Indeed, our analysis indicates that the encoder of body movements and the prior network can cooperate and disentangle clearly the representation of the music on instrument level. This combination  allows us to generate new music by sampling discrete latent vectors from the trained prior distribution and passing them through the decoder of VQ-VAE. In addition to music generation, we also study the production of the exact content of music played in a video by introducing a content encoder used during the training of the prior network. The content would be additional characteristic of music such as Midi.  
\\
In summary, our main contributions are as follows: 
(i) We introduce the Multi-Instrumentalist Net which is the first unsupervised system that for a given video of a musician playing an instrument from a variety of 13 instruments, will generate the associated music for the instrument in the video. We evaluate MI Net on videos and music in URMP dataset~\cite{li2018creating} recorded in a studio setting and demonstrate that MI Net can generate music with a reasonable quality and specific to the instrument being played. (ii) We demonstrate that the introduction of an additional content condition to the prior network can generate the corresponding music piece being played in the video. (iii) We further evaluate the MI Net on `in the wild' instrumental performance videos and discuss potential future directions. 

\section{Related Work}

In our work we propose to use the visual information of a musician playing an instrument to generate the music that captures the specificity of the video and audio contents. To set up the framework, we describe here related work in music generation and audio-visual tasks.

\textbf{\textit{Music Generation.}} Several deep learning methods have been introduced to generate novel music. In particular, autoregressive models that generate audio waveforms directly, such as Wavenet~\cite{oord2016wavenet}, SampleRNN~\cite{mehri2016samplernn}, and their variants~\cite{oord2018parallel,ping2018clarinet,dieleman2018challenge} have been shown to be successful in generation of speech or music signals. Since capturing high-level structure in audio waveforms is challenging, methods such as GANSynth~\cite{engel2019gansynth} and MelNet~\cite{vasquez2019melnet} proposed to use time-frequency representations, e.g., a spectrogram, to learn the generative model. 
Recently, non-autoregressive models such as MelGAN~\cite{kumar2019melgan} demonstrated convincing results on audio generation. In addition to spectrogram, symbolic musical representation (Midi) has been found instrumental in modeling and generating music~\cite{huang2018music,hawthorne2018enabling}. While the aforementioned methods generate music that is unconditional, there has been progress in generation of music with constraints. For example, it was proposed to constrain generative models to sample with respect to some predefined attributes~\cite{engel2017latent}. The Universal Music Translation network aims to translate music across various styles via raw audio waveforms. Works such as Jukebox~\cite{dhariwal2020jukebox} and MuseNet~\cite{payne2019musenet} showed the possibility to generate music based on user preferences, which translate to network model specifically trained with labeled tokens as a conditioning input.
Furthermore, recently, the Transformer autoencoder have been proposed to aggregate encoding of the Midi data across time to obtain a global representation of style from a given performance. Such a global representation can be used to control the style of the music~\cite{choi2019encoding}.
\\
\textbf{\textit{Audio-visual learning.}} While numerous methods have been introduced to work with the sound signal and its various representations, there is a possibility to add additional information that can help enhance the audio signal interpretation and generation. Indeed, the field of \textit{Audio-visual learning} deals with exploration and leveraging of the correlation of both audio and video for tasks that simultaneously involve these two signals. In recent years, methods for audio-visual learning have gained significant development and unlocked novel applications. For example, conditioning the visual and the sound streams on each other as a training supervision was shown as an effective training method for networks with unlabeled data in the audio-visual correspondence task~\cite{arandjelovic2017look,aytar2016soundnet,harwath2016unsupervised,owens2016ambient}. Moreover, it was shown that it is possible to separate object sounds by inspecting the video cues of an unlabeled video sound separation~\cite{gao2018learning,zhao2018sound,zhao2019sound,gan2020music}, or performing an audio-visual event localization task on unconstrained videos~\cite{tian2018audio}.  
\\
\textbf{\textit{Audio to Video Systems.}} Transformations between audio and video have been studied as well. In the audio-to-video direction, deep learning RNN based strategies were proposed to generate body dynamics correlated with sounds from audio-stream~\cite{shlizerman2018audio,ginosar2019learning}. Moreover, systems that generate parts of the face or synchronize lips movements from speech audio were shown to be possible~\cite{suwajanakorn2017synthesizing,jamaludin2019you,oh2019speech2face}.
\\
\textbf{\textit{Sound Generation from Videos.}} The direction of generating sound from video is a challenging problem. Initial deep learning work~\cite{owens2016visually} implemented a recurrent neural network to predict the impact sound features from videos and then was able to produce a waveform from these features. Later, a conditional generative adversarial network~\cite{chen2017deep} was proposed to achieve cross-modal audio-visual generation of musical performances. A single image is used as an input and the network performs supervision on instrument classes to generate a low-resolution spectrogram. In addition, a SampleRNN-based~\cite{zhou2018visual} have been introduced to generate natural sounds, e.g., baby crying, water flowing, given a visual scene. Later, an audio forwarding regularizer that considers the real sound as an input and outputs bottle-necked sound features showed that it can provide stronger supervision for the natural sound prediction and to produce associated sounds only from visual features~\cite{chen2020generating}.\\
Compared to natural sounds that have relatively simple structure, music across different instruments contains more complex elements. Previous work on music generation from videos was mostly focused on piano performance. A ResNet-based method was proposed to predict the pitch and the onsets events given piano video frames stream~\cite{koepke2020sight}. Audeo~\cite{su2020audeo} succeeded to transcribe a silent piano performance video to high-precision audio outputs. While the results of such methods are promising, the generation is limited to piano only. For woodwinds and brass instruments, it is unlikely to transcribe the music only from the visual stream since changes of the air blown into the instrument can produce different pitches with very minor visual changes. Recently, Foley Music~\cite{gan2020foley} proposed a Graph-Transformer network to generate Midi events from body keypoints and achieved convincing and robust outcomes. However, it includes limitations of using a different model per instrument, and it requires instrument labels to synthesize Midi events. In comparison, our method can generate different instruments music with the training process being completely unsupervised.
\begin{figure*}[t]
    \centering
    \includegraphics[width=\linewidth,height=3.5cm]{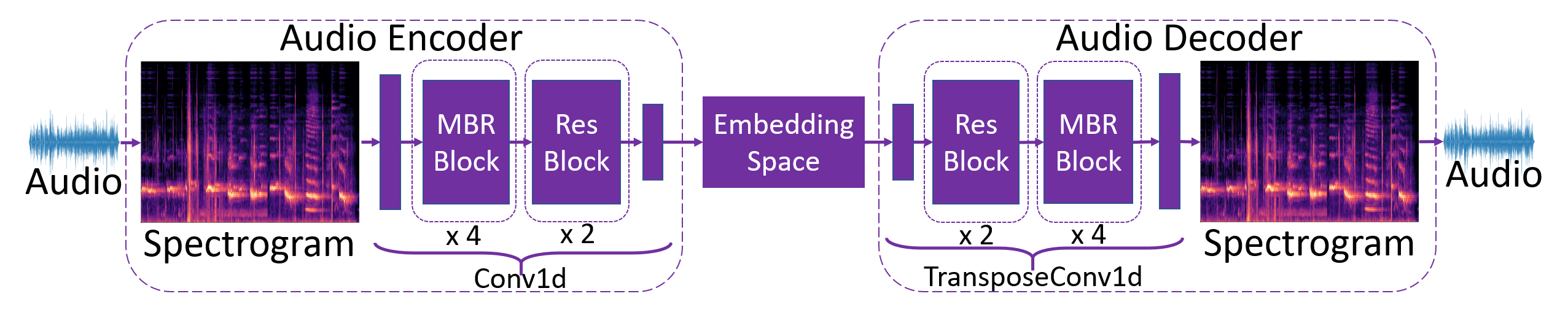}
    \caption{Detailed schematics of the components in VQ-VAE. The encoder and the decoder contain multiple multi-band residual (MBR) blocks and common residual blocks to reconstruct the input audio.}
    \label{fig:vq-vae}
\end{figure*}
\section{Methods}
\textbf{\textit{Visual Representations.}} We use human pose keypoints to capture cues of body motion that express playing an instrument. We use the OpenPose framework~\cite{cao2018openpose} to detect body and hand key points from each video frame and then stack the $2$D coordinates over time to structured visual representations (matrices). In practice, we find that the upper body keypoints are sufficient, which results in $5$ keypoints for the body parts and $21$ keypoints for each hand in total. To remove noise, we perform a linear interpolation of missing frames. Specifically, the joints that were not predicted well are interpolated linearly according to the distance to the previous and post detected frames. This prediction is based on the relative position of the joints to the precedent joint and ensures stability in the absolute position.\\
\textbf{\textit{Audio Representations.}} The choice of the correct audio representation is key in learning a generative music model. The straightforward representation is the audio waveform, however, training on waveform signals is challenging as described in previous works~\cite{vasquez2019melnet} and would typically take a long time to converge to the desired performance. Another common representation is symbolic Midi. While informative, Midi is not applicable in our case since we aim to design an unsupervised model suitable for different instruments. Midi explicitly uses the instrument name to synthesize the audio via a Midi synthesizer. Alternatively, frequency representation could be used for audio representation. We use the magnitude of log-spectrogram as the audio representation by applying Short-Time Fourier Transform to the waveform resulting in a $2$D time-frequency representation $S = F\times T$, where $F$ is the number of frequency bins and $T$ is the number of time steps. We learn the latent representation of the spectrogram via a reconstruction task described in the following sections.
\subsection{Encoding of Audio Features}
\textbf{\textit{VQ-VAE.}} To encode the magnitude of the log-spectrogram into the latent space, we introduce a multi-band residual 1D convolutional Vector Quantized Variational Autoencoder (VQ-VAE), as shown in Fig.~\ref{fig:vq-vae}. VQ-VAE~\cite{van2017neural} is a type of VAE~\cite{kingma2013auto} in which the encoder outputs a discrete latent representation. The decoder decodes this representation and reconstructs the input. The prior in this network is being learned rather than being static. VQ-VAE have been shown to successfully learn latent representations utilized for generation of high-quality images, videos, and audio. In our case, the audio encoder network encodes the log-spectrogram to a discrete latent representation, and the audio decoder decodes it to reconstruct the log-spectrogram. In general, we define a latent embedding space $e\in \mathbb{R}^{K\times D}$ where $K$ is the size of the discrete latent space such that it is a $K$-way categorical, and $D$ is the dimension of each latent embedding vector $e_i$. During forward propagation, the continuous representation encoded by the audio encoder is replaced with its closest discrete vector. This is defined as follows, let $Z = E(S)\in \mathbb{R}^{D\times T}$ be the output of the audio encoder before quantization. For each time step $t$, VQ-VAE finds the nearest vector in the codebook and uses it as the latent representation, i.e. $\text{Quantize}(Z_t)=e_k$ where $k = \arg \min_i \Vert Z_t-e_i\Vert^2_2$. The vector $e_k$ is then passed to the decoder to decode and reconstruct the log-spectrogram $S$. The VQ-VAE model incorporates two additional terms in its objective to align the vector space of the code with the output of the encoder. (i) The embedding loss is applied to the codebook variable and $e_k$ and brings the selected codebook $\mathbf{e}$ closer to the output of the encoder $E(S)$. (ii) The commitment loss is applied to the encoder weights which aims to keep the output of the encoder as close as possible to the chosen codebook vector to prevent it from fluctuating from one code vector to another. As proposed in ~\cite{van2017neural}, we use the exponential moving average updates for the codebook as a replacement for the embedding loss. We define the resulting loss as
$
    \mathcal{L} = \Vert S - D(e)\Vert^2_2 + \beta \Vert E(S)-sg[e]\Vert^2_2,
$
where the first term is the log-spectrogram reconstruction loss and the second term, $\beta$, is a hyper-parameter which depends on the scale of the reconstruction loss and $sg$ stands for the stop gradient operator defined as the identity at forward computation time and has zero partial derivatives. Since we assume a uniform prior for the latent space, the KL term that appears in the ELBO is constant with respect to the encoder parameters and is not included.\\ 
\textbf{\textit{Multi-band Residual Blocks.}} The log-spectrogram is of high dimension due to the desired high resolution on the frequency bins. Inspired by the PerformanceNet~\cite{wang2019performancenet}, we thereby use a multi-band residual learning method on the audio encoder and the decoder to better capture the spectral features of musical overtones. The multi-band residual (MBR) block splits the input into a specific number of frequency bands and then feeds each band individually to identical sub-blocks consisting of the following layers: 1D-convolution, ReLU, 1D-convolution. The output of all sub-blocks are then concatenated along the frequency dimension and a residual connection sums up the output with the input of the block. In the audio encoder, we progressively divide the spectrogram into more bands in the earlier layers, and into fewer bands in the latter layers. The decoder then decodes the latent representation from fewer bands to more bands in a symmetric way. In our proposed VQ-VAE architecture, the audio encoder receives the log-spectrogram as an input and passes it through a 1D convolutional layer. The features then go through four MBR blocks described above. Subsequently, it passes through two 1D convolutional residual blocks and a single 1D convolution to generate the continuous latent representation which is mapped to the discrete latent space. The resulting discrete latent features are then fed into the audio decoder which is a mirrored structure of the encoder. Since both the encoder and the decoder are all fully 1D convolutional, the log-spectrogram input generally supports any length of input.
\subsection{Encoding Visual Features and Learning a Prior over the Audio Latent Code}
\begin{figure}[t]
    \centering
    \includegraphics[width=0.6\linewidth]{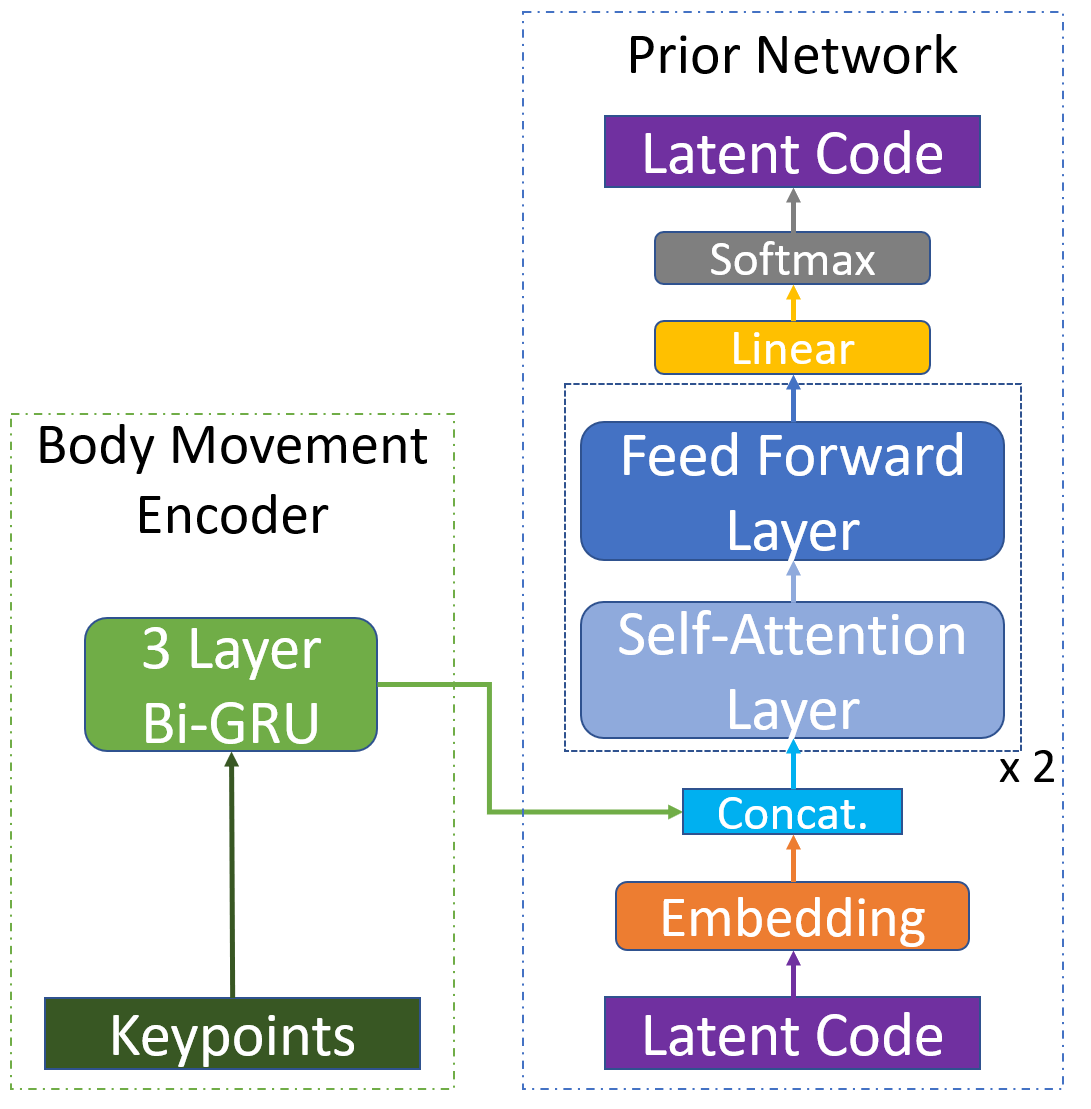}
    \caption{Detailed schematics of the components in the Body Movement Encoder and the Prior Network.}
    \label{fig:uncond_prior}
\end{figure}
\textbf{\textit{Body keypoints encoder.}} Given a sequence of 2D human pose key points $P = \{p_1, p_2,...,p_T\}\in \mathbb{R}^{J\times T}$, where $J$ is the number of joints and $T$ is the total number of time steps, we first encode them into a latent representation. This latent space should differentiate movements of playing different instruments and self-organize itself into separate clusters as shown in the recent unsupervised skeleton-based action recognition approach~\cite{zheng2018unsupervised, su2020predict}. To achieve that, we use a bidirectional Gated Recurrent Units (GRU) as the body movement encoder $E_b$. Given the input sequence $P$, the last hidden state of the encoder $h = E_b(P)$ can be seen as the global representation of the musician's movement. In order to associate the body movements with audio features, we \textit{jointly train the body keypoints encoder} with \textit{the prior of the latent space of the audio features}. Superimposed with the latent audio features, the body movements features differentiate the type of instrument performance and other characteristics of the performed music. This allow us to use body movements to generate new music of the corresponding instrument.\\
\textbf{Learning a Prior over the Latent Space.} The prior distribution over the discrete latent audio features $p(Z)$ is a joint distribution of categorical across time and can be learned autoregressively. When training the VQ-VAE, the prior is kept constant and uniform. After training, we fit an autoregressive distribution over $Z$ so that we can generate new samples via ancestral sampling as shown in Fig.\ref{fig:uncond_prior}. We use the encoder of transformer structure similar to the GPT-1~\cite{radford2018improving} over the discrete latent space. We concatenate the last hidden state of the body keypoints encoder $h$ to every time steps of the discrete latent representation. This forces the prior of audio features to align and correlate to those body motion features when autoregressively learning to predict the next latent code. Subsequently, the concatenated features are passed through the multi-head self-attention layer of the scale dot-product self-attention defined as: Attention$(Q,K,V) =$softmax$(\frac{QK^T}{\sqrt{D_k}})V$,
where $Q,K$, and $V$ are query, key and value, respectively. The layer calculates a weight by dot products of the key $K$ and query $Q$, and then outputs a weighted sum of the value $V$. Using multi-head self-attention allows the model to integrate information from different independent representations. Next, the point-wise feed-forward layer takes the input from the self-attention layer, and further transforms it through two fully connected layers with ReLU activation as: Feed Forward$ = \max(0,xW_1 + b_1)W_2 + b_2.$ The outputs of the self-attention and feed forward layers are passed to a softmax layer to predict the probability of the next latent code over the codebook.
\subsection{Content Conditioning}
\begin{figure}[t]
    \centering
    \includegraphics[width=0.8\linewidth]{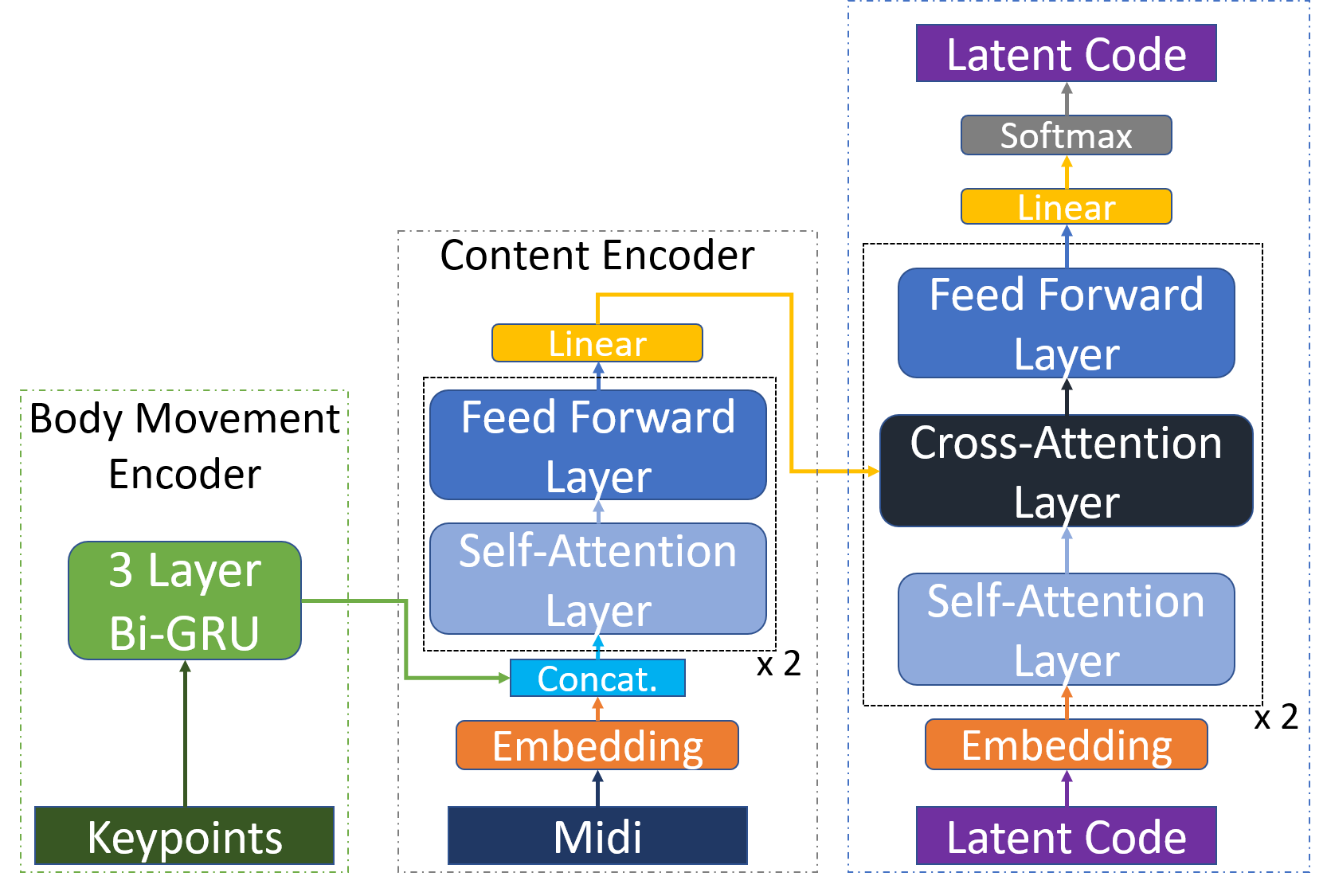}
    \caption{detailed schematics of the components in prior network with content conditioning.}
    \label{fig:cond_prior}
\end{figure}
In addition to the unconditional generation of music, it is possible to generate actual music content of the video. Therefore, we also explore whether we can add content conditioning the prior network and to generate the exact music content for a specific instrument. We use the Midi as the content signal. The Midi is considered as a matrix $M \in N\times T$ where $N$ is the number of notes and $T$ is the number of time steps. Since all considered instruments in our experiments are monophonic, there is at most one active note at each time step. We first convert the 2D matrix into a binary matrix by ignoring the expressive dynamics (i.e., the loudness of music). We then transform the binary matrix to a 1D sequence that contains the index of the activated note at each time step. We use a transformer-based encoder-decoder architecture~\cite{vaswani2017attention} to achieve the content conditioning. The content encoder is the transformer encoder that takes the Midi information as input as shown in Fig.\ref{fig:cond_prior}. In this case, we concatenate the encoded body movement representation to the embedded Midi at each time step and pass them to two self-attention and Feed-forward layers. The content encoder output will go through a fully connected layer such that it becomes the conditioned signal $C$. The transformer decoder is similar to the unconditional prior network except that we add a cross-attention layer after the self-attention layer to compute the attention between the conditioned signal and the latent code representation. Considering the output of the self-attention module $A\in \mathbb{R}^{T_a \times C}$ and the Midi conditional signal $M\in \mathbb{R}^{T_m \times C}$, where $C$ is the feature dimensions, the cross-attention is defined as: Cross Attention$(A,M) =$ softmax$(\frac{AM^T}{\sqrt{D_k}})M.$ Here, the feature dimension of $A$ and $M$ are designed to be the same. During sampling, we provide the Midi content and body movements of single instrument performance to the body movement encoder and to the content encoder, respectively. The prior network then autoregressively generates discrete latent representations via ancestral sampling. The discrete latent representations will feed to the decoder of VQ-VAE and generate the instrument's music associated with the exact instrument and content in the video.
\section{Experiments \& Results}
\textbf{Datasets.} We evaluate the MI Net on \textbf{URMP} dataset~\cite{li2018creating}, a high-quality multi-instrument video dataset recorded in a studio. It includes $13$ instruments and provides the musical score in the Midi format which we use for evaluation. There are $44$ videos and $148$ tracks in total. We use $135$ tracks for training and $13$ tracks for testing such that each instrument has at least one track in the test set. We further evaluate the MI Net on \textbf{Solos} dataset~\cite{montesinos2020solos}, a very recently published dataset of YouTube videos containing excerpts of musicians playing different instruments for auditions. It contains the same $13$ instruments as the URMP dataset. There are $755$ videos in total, and each instrument has approximately $58$ videos on average. We use a 9:1 ratio to split the training and the testing sets. Solos also contains pre-processed skeleton keypoints extracted via Openpose however doesn't include Midi files due to `in the wild' nature of the videos.\\
\textbf{Implementation details.} We use Pytorch to implement our MI Net. The sampling rate of all audios is set as in 16Khz, and we use 1024 frequency bins with a hop size of 256 to generate the log-spectrograms. In training, we randomly select 4 seconds segments from videos. The VQ-VAE, includes 1D convolutions in all MBR and residual blocks whose outputs are of 512 channels. We reduce the audio encoder outputs to 64 channels and map it to an embedding space with the size of 1024. The body movement encoder is a 3-layer bi-GRU with hidden size of 32. For all self-attention modules, we use 2 layers of 128 dimensions and 8 heads. More details of architectural design can be found in the Supplementary material.\\
\textbf{Comparison with Other Models.} To our best knowledge, we are the first work in the direction of unsupervised generation of instrument specific music. Therefore, we additionally implement two baselines for comparison. (i) \textbf{RNN-based Seq2Seq Network:} We implement an encoder-decoder recurrent neural network. The encoder takes body keypoints as input, and the decoder generates the expected spectrogram. (ii) \textbf{Graph-Transformer Network:} We implement a Graph-Transformer network similar to the architecture of Foley Music~\cite{gan2020foley}. It is based on Spatio-temporal Graph Convolution Network (ST-GCN)~\cite{yan2018spatial} which encodes the body keypoints to pose features, which are then fed into the Transformer decoder where each block contains the self-attention, cross-attention, and feed-forward modules. Instead of predicting the Midi events, we directly generate the log-spectrograms.\\
\textbf{Evaluation.} Evaluation of generative models is not a well defined procedure, particularly for MI Net which aims to generate perceptually-realistic
audio. We thereby evaluate our method against a diverse set of metrics, each of which captures a particular aspect of the model performance.\\
\begin{figure}[t]
    \centering
    \includegraphics[width=0.9\linewidth]{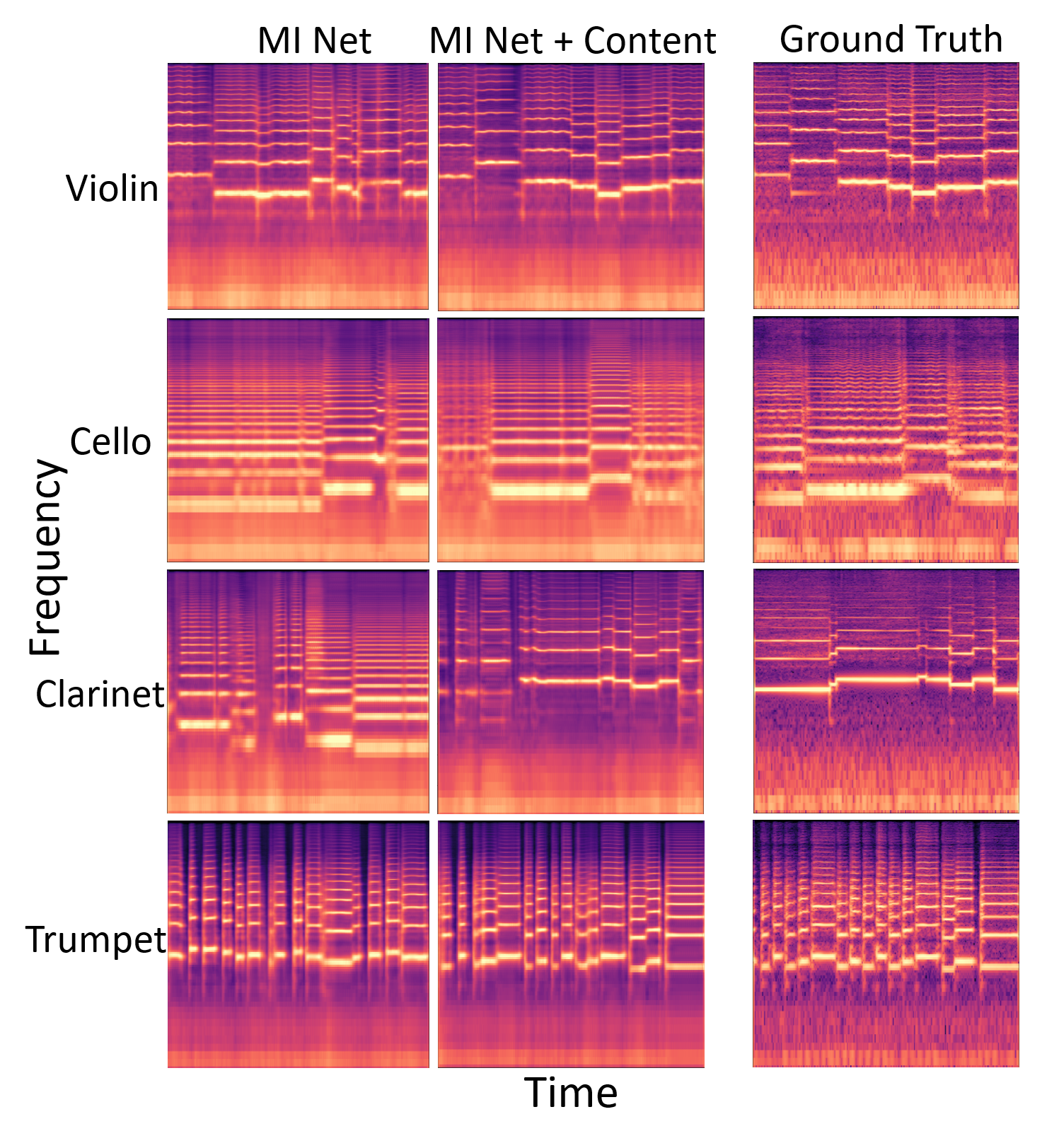}
    \caption{Examples of generated spectrograms for four instruments in URMP dataset. Left: Generated samples without content condition. Middle: Generated samples with content condition. Right: Ground Truth.}
    \label{fig:spec}
\end{figure}
\textbf{1) Number of Statistically-Different Bins (NDB).} We adopt the metric proposed in ~\cite{richardson2018gans} and used in ~\cite{engel2019gansynth, gan2020foley} to measure the diversity of the generated examples: first, the training examples are clustered into $k=50$ Voronoi cells by k-means in log-spectrogram space. The generated examples are also mapped into the same space and are assigned to the nearest cell. NDB is reported as the \textit{number of cells where the number of training examples is statistically significantly different from the number of generated examples by a two-sample Binomial test.} For each model, we generate about $1600$ samples from the testing set and perform the comparison. For a reference, we also evaluate the NDB on the testing data itself. The NDB results are shown in Table \ref{tab:ndb}. NDB indicates how well model learns from the training set. Here, the larger the number is (up to 50), the less similar generated samples are compared to the training set, which means worse learning performance. For the URMP dataset, the MI Net outperforms other methods by a large margin. Both the RNN-based Seq2Seq and Graph-Transformer do not generate music with sufficient quality under the unsupervised setup. In addition, the generated samples of MI Net without conditioning on content have a distribution closer to the training data, therefor, even resulting in a lower NDB score than the reference. Once the content condition is added, the distribution of the generated samples becomes closer to testing set. For Solos dataset, while the NDB result of our method is better than others, the generated samples are not satisfactory. One of the issues is the body motions of `in the wild' videos have large variations and becomes challenging to differentiate in the unsupervised manner as we analyze in the below metric.\\
\begin{table}[]
\small
\centering
\caption{NDB results. \textbf{Lower is better}.}
\label{tab:ndb}
\begin{tabular}{|c|c|c|}
\hline
Model                           & URMP & Solos \\ \hline
RNN-based Seq2Seq               &   48 & 44 \\ \hline
Graph-Transformer               &   45 & 41 \\ \hline
\textbf{MI Net (Our)} &   \textbf{33} & \textbf{36} \\ \hline
\textbf{Content Cond. MI Net (Our)} &   \textbf{35} & - \\ \hline
Testing Set Data &  36 & 31 \\ \hline
\end{tabular}
\end{table}
\begin{figure*}[h]
    \centering
    \includegraphics[width=0.7\linewidth]{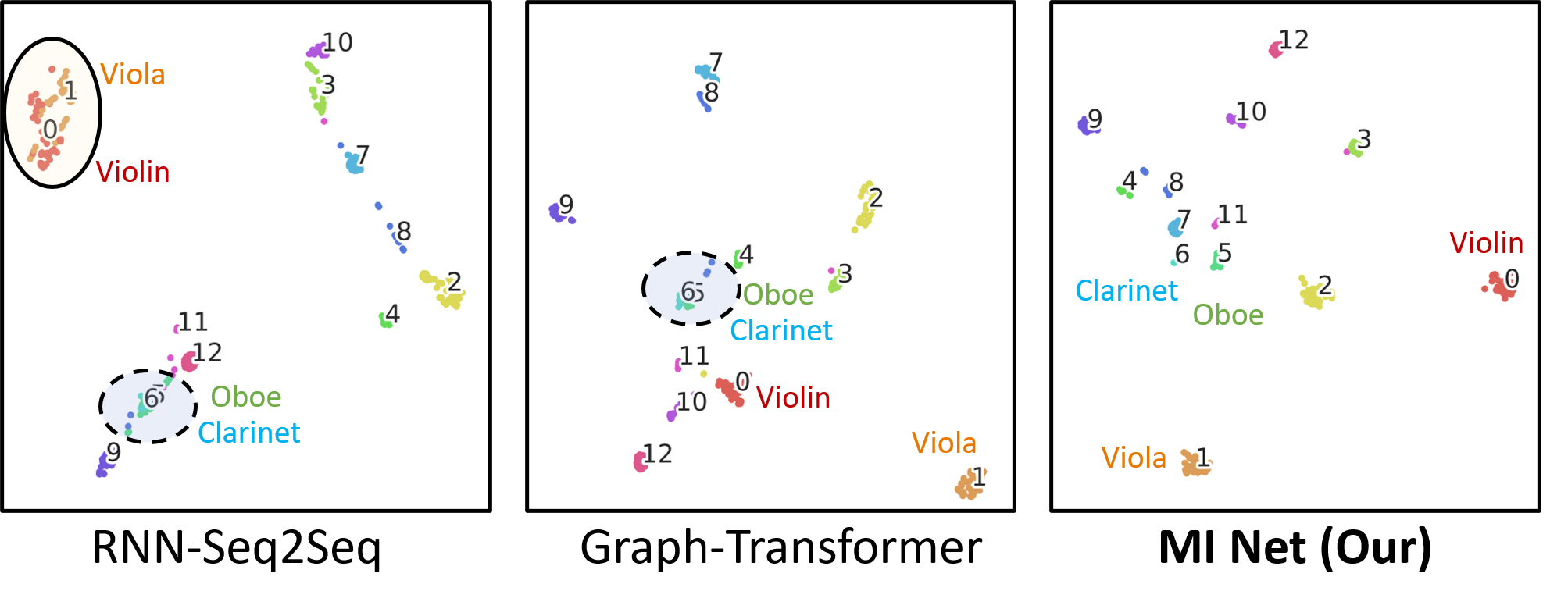}
    \caption{T-SNE plots of the encoded body movements representation in URMP test set. the number next to clusters indicates instruments: 0. Violin, 1. Viola, 2. Cello, 3. Double bass, 4. Flute, 5. Oboe, 6. Clarinet, 7. Bassoon, 8. Saxophone, 9. Trumpet, 10. Horn, 11. Trombone, 12. Tuba. Dotted circle: Mixing of samples belonging to Oboe and Clarinet instruments. Solid circle: Mixing of samples belonging to Violin and Viola instruments.}
    \label{fig:tsne}
\end{figure*}
\textbf{2) Classification with Body Motion Features.} To evaluate whether the encoded pose features are separated to generate exclusive music for specific instruments, we extract the body movement encoder's final hidden state and fit a K-Nearest-Neighbors classifier ($K=1$) using cosine similarity metric. For comparison, we use the encoder final state for RNN-based Seq2Seq, and take the mean of both temporal and joints dimensions of the outputs of ST-GCN for Graph-Transformer to perform the classification. The results are shown in Table~\ref{tab:knn}. Notably, for URMP dataset, our method associates the body movements with the instruments at a high accuracy, evident by the score in Table~\ref{tab:knn} and t-SNE plots of the latent representations in Fig.\ref{fig:tsne}. While the results of RNN-Seq2Seq and Graph-Transformer are of reasonable scores, the plots show that these models do not precisely distinguish the instruments. Furthermore, they cannot generate expected instrumental music since these models do not find the correlations between audio features and body movements but rather separate the instruments according to the difference of poses only. This becomes challenging for instruments with similar poses (e.g., Viola v.s. Violin, Oboe v.s. Clarinet). In comparison, our method learns the mapping of the latent spaces, containing more representative information to allow the model builds the connection between visual and audio events. As expected, in the more challenging `in the wild' dataset of Solos, our method outperforms other methods significantly. However, we realize that the $61.5\%$ accuracy is not enough for MI Net to fully distinguish the body movements and fully generate satisfactory music. We describe the limitations and possible improvements in Section 5.\\
\begin{table}[]
\small
\centering
\caption{KNN Classification Accuracy in $\%$.}
\label{tab:knn}
\begin{tabular}{|c|c|c|}
\hline
Model                           & URMP  & Solos \\ \hline
RNN-based Seq2Seq               &  82.6  &  37.8\\ \hline
Graph-Transformer               &   89.2 &  38.3\\ \hline
\textbf{MI Net (Our)} &   \textbf{93.7} & \textbf{61.5} \\ \hline
\end{tabular}
\end{table}
\textbf{3) Qualitative Human Evaluation}
\begin{table}[]
\small
\centering
\caption{Human evaluation of real vs. fake audio samples. Success means the percentage of the generated sound by MI Net that was considered real (out of $50\%$ selected by the Oracle).}
\label{tab:real-fake}
\begin{tabular}{|c|c|c|}
\hline
Method                           & \textbf{MI Net (Our)} & Oracle \\ \hline
Success Rate &     24     & 50     \\ \hline
\end{tabular}
\end{table}
We evaluate the quality of generated samples by performing a human evaluation. We provide real (which originally belongs to the video) and fake (MI Net) audio to the Amazon Mechanical Turk (AMT turkers). The turkers are asked to choose the audio that they believe is real. We surveyed 50 participants individually, where each participant evaluated 39 pairs of 4 seconds audios (3 samples per instrument). To be noted, an oracle score
of 50\% indicates perfect confusions between real and fake. Since the RNN-based Seq2Seq and Graph-Transformer cannot generate music compared with ground truth, we evaluate the MI Net on URMP dataset. The result in Table~\ref{tab:real-fake} shows that our generated samples could fool the participants at a reasonable success rate of 24\%.\\
\textbf{Ablation Study.} We perform the ablation studies to evaluate the impact of each component of our method. We use URMP dataset to perform the comparison experiments.\\
\textbf{1) MBR.} In VQ-VAE, we utilize the multi-band residual blocks (MBR) to learn the representation from a high-resolution log-spectrogram. We compare our method with common 1D convolutional without MBR to show the effectiveness. We compare the testing L2 reconstruction loss of spectrogram, and the results in Table~\ref{tab:ablation-vqvae} shows the benefit of MBR blocks. \\
\begin{table}[]
\small
\centering
\caption{Ablation study of the architectural design of VQ-VAE in terms of reconstruction (L2) loss.}
\label{tab:ablation-vqvae}
\begin{tabular}{|c|c|}
\hline
VQ-VAE                        & L2 Loss \\ \hline
1D Conv w.o. MBR              &     0.89\\ \hline
\textbf{1D Conv w. MBR (Our)} &     \textbf{0.78}\\ \hline
\end{tabular}
\end{table}
\begin{table}[]
\small
\centering
\caption{Ablation study of the architectural design of the Content Condition in terms of NELL loss computed for the latent code prediction.}
\label{tab:ablation-midi}
\begin{tabular}{|c|c|}
\hline
Midi Cond. Structure       & NLL Loss \\ \hline
GRU-Seq2Seq                &    2.93\\ \hline
\textbf{Content Encoder + Prior Net (Our)} &    \textbf{2.55}\\ \hline
\end{tabular}
\end{table}
\textbf{2) Content Encoder.} We use a transformer-based architecture to learn the Prior with the content condition. To verify its effectiveness, we replace the Transformer-based architecture with a GRU-seq2seq model. As shown in Table \ref{tab:ablation-midi}, our method achieves lower negative-log likelihood loss than the baseline. This demonstrates the benefit of our design choice to capture the dependencies between content and latent audio representations.
\section{Limitations}
Our experimental results show that MI-Net can generate instrumental music from unlabeled videos recorded in a studio. However, for videos `in the wild', generation of quality music remains to be challenging. We identify two main challenges that limit the performance. (i) For a dataset with a large variance in music, the current VQ-VAE model cannot reconstruct the spectrogram with resolution due to a limited latent space. (ii) Variations in views and subjects make it harder to differentiate instruments in an unsupervised way. A more Specifically designed body movement encoder would be required. These two challenges are interconnected since the latent spaces are interconnected. If one of the latent spaces is not learned well, it would be hard to generate satisfactory music.
\section{Conclusion}
We propose an unsupervised system named Multi-Instrumentalist Net (MI Net) that generates the associated sound for a video via human body movement playing an instrument. We demonstrate that the MI Net can generate reasonable quality music on the URMP dataset recorded in a studio setting. Besides, we evaluate the MI Net on `in the wild' videos and discuss the limitations and potential future research directions.

{\small
\bibliographystyle{ieee_fullname}

}

\end{document}